# COMPUTATION OF THE SPATIAL STRING TENSION IN HIGH TEMPERATURE SU(2) GAUGE THEORY[*]

G.S. BALI, K. SCHILLING,
*Fachbereich Physik, Universität-Gesamthochschule Wuppertal, Gausstr. 20*
*D-42097 Wuppertal, Germany*

J. FINGBERG, U.M. HELLER
*SCRI, The Florida State University*
*Tallahassee, FL 32306-44052, U.S.A.*

and

F. KARSCH
*HLRZ, c/o KFA Jülich*
*D-52425 Jülich, Germany*
*and*
*Fakultät für Physik, Univ. Bielefeld, P.O. Box 100131*
*D-33501 Bielefeld, Germany*

A detailed investigation of the temperature dependence of the spatial string tension $\sigma_s$ in $SU(2)$ gauge theory is presented. A sustained performance of 3 GFLOPS on a 64K Connection Machine CM-2 equivalent has been achieved. Scaling of $\sigma_s$ between $\beta = 2.5115$ and $\beta = 2.74$, on large lattices, is demonstrated. Below the critical temperature, $T_c$, $\sigma_s$ remains constant. For temperatures larger than $2T_c$ the temperature dependence can be parametrized by $\sigma_s(T) = (0.369 \pm 0.014)^2 g^4(T) T^2$, where $g(T)$ is a 2-loop running coupling constant with the scale parameter determined as $\Lambda_T = (0.076 \pm 0.013) T_c$.

## 1. Outline

After a brief introduction into lattice gauge theories where we concentrate on the aspects of computational relevance, we discuss our implementation of the updating algorithm on the Connection Machine CM-2. This is followed by a presentation of the physical background, motivation, and a summary of related results, obtained by previous authors (Section 3). In Section 4 technical aspects of our data evaluation

---

[*]Talk presented by Gunnar Bali at the "Workshop on Large Scale Computational Physics on Massively Parallel Computers", HLRZ, Jülich, June 14-16, 1993.

procedure are discussed. The paper concludes with physical results (Section 5).

## 2. Simulation of SU(2) Gauge Theory

### 2.1. *The "mathematical" problem*

The lattice regularization is the only known non-perturbative method to treat quantum field theories quantitatively. A space-time lattice with lattice spacing $a$ and volume $(N_\sigma^3 \times N_\tau)a^4$ is used. Physical observables can be extracted in the continuum limit $(a \to 0)$. An element of the gauge group, responsible for interactions, is mapped to each link connecting two lattice sites. In the physical theory of interest, the theory of strong interactions, QCD, this group is $SU(3)$. Because of the similarity of $SU(3)$ vacuum structure with the one of $SU(2)$ we simulate $SU(2)$, in order to minimize the computational task.

A set of group elements (link variables) $\mathcal{C} = \{U_\mu(n)\}$ with $\mu \in \{1,\ldots,4\}, n = (n_1, n_2, n_3, n_4)$, $n_i \in \{1,\ldots,N_\sigma\}$, $n_4 \in \{1,\ldots,N_\tau\}$, and $U_\mu(n) \in SU(2)$ is called a configuration. Each configuration represents a possible field distribution on the finite lattice. In the path integral formulation of quantum field theories, expectation values of operators $\langle O \rangle$ (observables) can be calculated as an average of the values $O$ takes on all possible configurations, weighted by a Boltzmann measure:

$$\langle O \rangle = \frac{1}{Z} \int \left( \prod_{n,\mu} dU_\mu(n) \right) e^{-\beta S(U)} O(U) \quad , \quad Z = \int \left( \prod_{n,\mu} dU_\mu(n) \right) e^{-\beta S(U)} \quad , \tag{1}$$

with $\beta S$ being a discretized version of the continuum action of the Euclidean theory.

The simplest possible gauge invariant choice with the desired continuum limit is the so-called Wilson action:

$$S = \sum_{n,\mu > \nu} \left( 1 - \frac{1}{2} \operatorname{Tr} U_{\mu\nu}(n) \right) \quad , \tag{2}$$

where $U_{\mu\nu}(n)$ denotes the ordered product of all link variables around the elementary square (plaquette) emanating from site $n$ into the positive oriented $\mu\nu$-plane. In this choice $\beta = 4/g^2$ with $g$ being the bare (lattice) coupling parameter. By changing $\beta$ one can control the lattice spacing $a$. Since nonabelian gauge theories are asymptotically free, the continuum limit corresponds to infinite $\beta$. The dimensionful quantity $a$ can me measured in units of some physical scale, e.g. a mass like the square rott of the string tension $\sqrt{\sigma}$, or the critical temperature $T_c$.

The (high dimensional) integral Eq. 1 can be calculated numerically by use of stochastic (Monte-Carlo) methods. As most possible configurations are strongly suppressed by their Boltzmann weight factors, one better concentrates on the *essential* configurations. This is achieved by the technique of importance sampling: Configurations are generated according to the probability density

$$dP(\mathcal{C}) = \left( \prod_{n,\mu} dU_\mu(n) \right) e^{-\beta S(\mathcal{C})} \quad . \tag{3}$$

After generation of a suitable number of such (independent) configurations the observable of interest is calculated as an average of the values of the corresponding operator on the ensemble of generated gauge fields.

The ensemble is produced by a Markov process: Each configuration only depends on a finite number of preceding configurations: $\mathcal{C}_n = f(\mathcal{C}_{n-1}, \mathcal{C}_{n-2}, \ldots, \mathcal{C}_{n-m})$ with $m$ being the order of the Markov-chain. A criterion which guarantees the right equilibrium distribution for the Monte-Carlo simulation is the so-called detailed balance condition

$$dP(\mathcal{C}_{n-1})W(\mathcal{C}_{n-1} \to \mathcal{C}_n) = dP(\mathcal{C}_n)W(\mathcal{C}_n \to \mathcal{C}_{n-1}) \tag{4}$$

with $W(\mathcal{C}_{n-1} \to \mathcal{C}_n)$ being the appropriate transition probablility. In addition to this requirement the algorithm has to be ergodic, i.e. there has to be a nonvanishing probability for each possible configuration to be generated within a finite number of steps. For a limited simulation time this means that the trajectories should cover the phase space as uniform as possible.

### 2.2. Implementation on the Connection Machine CM-2

There is some freedom in choosing $W(\mathcal{C}_{n-1} \to \mathcal{C}_n)$ according to Eq. 4. In our implementation we use a combination of two algorithms for updating the gauge fields. One of them is the heatbath algorithm, which allows a change in the action with transition probability

$$W(\mathcal{C}_{n-1} \to \mathcal{C}_n) \propto \exp(-\beta S(\mathcal{C}_n)) \quad . \tag{5}$$

The necessary group measure is generated by the method proposed by Kennedy and Pendleton,[1] which is suitable for SIMD architectures. This method is combined with the deterministic microcanonical Creutz-overrelaxation[2] that provides a faster movement of gauge fields in order to fight critical slowing down. Both algorithms are mixed stochastically to a hybrid form with the probability of heatbath sweeps being varied from 1/4 to 1/14, depending on $\beta$, $N_\tau$, and our experience.

Due to the locality of the Wilson action (Eq. 2), the implementation on a serial machine is straight forward: One changes all $4N_\sigma^3 N_\tau$ links one at a time, either in a fixed order or randomly. After each link has been updated once (in the average) one has carried out a *sweep*. On a parallel machine one has to be more careful since a whole set of links is updated at once. Because of the Markov condition links that are updated at the same time must not depend on each other. With the Wilson action this means that two links being updated at once must not share a common plaquette. The simplest way to provide this is a red/black division of the lattice. Lattice sites with even sum of coordinates are called *red*, the others *black*. One loops over red/black, and the direction $\mu$. All of these eight subsets include no links sharing a common plaquette.

We implement the algorithm on a data parallel programming model (CMFortran) which is the only possibility on SIMD architectures but also useful for many implementations on MIMD machines, such as the CM-5. Each CM-2 has some fixed number of *physical* processors $N_{PP}$. In order to allow portability of programs between Connection Machines with different $N_{PP}$, and to support scalability of codes, physical processors are hidden to the Fortran programmer behind the (software) concept of *virtual* processors (VPs): Each site of a *parallel* array corresponds to a VP. For the programmer all (parallel) commands are performed on all VPs at one and the same time. Internally, a certain number of VPs is mapped to one PP (therefore, $N_{VP} \geq N_{PP}$ to avoid latency of parts of the machine), and the looping over these VPs is automatically done by the compiler. For instance, VPs can be thought of as sites in a space-time lattice. Then, the canonical data structure for the configuration $\{U_\mu(n)\}$ would be: U(4,$N_\sigma$,$N_\sigma$,$N_\sigma$,$N_\tau$), where U is an $SU(2)$ matrix, with the first dimension being serial (denoting the direction $\mu$ of the link), and the other dimensions — being parallelized over — addressing the lattice sites $n$.

Implementation of the red/black mask means that the computations are only performed on half of the lattice volume at once. Thus, half of the machine would be inactive. In order to establish load balancing it is advantageous to declare U(4,2,$N_\sigma$,$N_\sigma$,$N_\sigma$,$N_\tau$/2), instead. The second index is serial and labels red/black sites. This reduces the number of virtual processors to half of the lattice volume and guarantees that all processors are active all the time.

On the Connection Machine, there are still two problems with this kind of data layout. One is excessive memory requirement: Although they are not needed all the time (VP looping), all temporary arrays (as they are needed for matrix multiplications etc.) have parallel dimension equal to $N_{VP}$, which is half the lattice volume. So each temporary array consumes one eighth of the space of the configuration itself. The second problem is communication: Communication between VPs belonging to the same PP is much faster than communication between different PPs. The overall speed of a communication step (like sending a link to the left neighbor) is determined by the speed of the slowest communication channel involved. In CMFortran for CM-2 the programmer has only little influence on the mapping of VPs to PPs. So a mixture between on- and off-chip communication in one step can easily occur, depending on lattice, and machine geometries. From this point of view it is wise to emphasize on-chip communication by adopting the machine size in such a way to the computational problem that the VP-ratio ($N_{VP}/N_{PP}$) is large.

One possibility to circumvent these problems is blocking the lattice into hypercubes (of $2^4$ lattice sites each) by hand: U(4,16,$N_\sigma$/2,$N_\sigma$/2,$N_\sigma$/2,$N_\tau$/2) The second index labels the address within each hypercube (lowest bit: first direction etc.). The serial loop is now over 64 instead of eight subsets. Thus, memory requirement for temporaries is reduced by a factor eight! For one half of all communications a high speed is guaranteed since the sites are located on the same VP and "communication" is programmed by hand. The disadvantages are higher complexity of Fortran code, limitation to lattices of even extents $N_\sigma$ and $N_\tau$, and a lower VP-ratio (by this factor eight).

For an 8K CM-2 (*slicewise* model) like the one in Wuppertal $N_{VP} \geq N_{PP} = 2^{10}$.

This translates into $2^{14}$ lattice sites. Saturation in the sense that the GFlops rate remains constant within 10% is reached for $2^{16}$ lattice sites (e.g. on a $32^3 \times 2$ lattice). On the 16K CM-2 (*fieldwise* model) at HLRZ $N_{PP} = 2^{14}$. For saturation the lattice must have at least $32^4$ sites. Using the above kind of algorithm we reach a sustained all-over speed (updating and measurement) of 1.5 GFlops/64K (*fieldwise* model) at the HLRZ or 3 GFlops/64K (*slicewise* model) at the Wuppertal and SCRI CM-2s. The latter rate has to be compared with 4.5 GFlops/64K in the $SU(3)$ case where the computation/communication ratio is larger. To reach the above performance the matrix multiplications have been programmed in assembler language.

## 3. Physical Background

An understanding of nonabelian gauge theories is important for the physics of the quark-gluon phase of matter to be studied in heavy ion collisions. The temperature $T$ corresponds to the inverse temporal extent of the lattice: $aT = 1/N_\tau$. Non-abelian $SU(N)$ gauge theories in (3+1)-dimensions are known to undergo a deconfining phase transition at high temperature. The physical string tension, characterizing the linear rise of the potential between static quark sources with distance, decreases with increasing temperature and vanishes above $T_c$. The potential becomes a Debye screened Coulomb potential in the high temperature phase. While the leading high temperature behaviour as well as the structure of the heavy quark potential for temperatures well above $T_c$ can be understood in terms of high temperature perturbation theory, it also is expected that non-perturbative effects like the generation of a magnetic mass term, $m_m \propto g^2(T)T$ ($g(T)$ is a temperature dependent running coupling "constant"), in the gluon propagator may influence the spectrum in the deconfined phase. These non-perturbative effects in the magnetic sector will manifest themselves in correlation functions for the spatial components of gauge fields.

(3+1)-dimensional renormalizable quantum field theories at high temperature, through dimensional reduction, can be reformulated as effective 3-dimensional theories, with the scale of the dimensionful couplings given in terms of the temperature.[3] In the case of $SU(N)$ the effective theory includes interactions with adjoint matter (Higgs) fields, emerging from the temporal components of the original gauge fields. In particular, for $SU(2)$ the effective action reads[4,5]

$$\beta_3 S_E = \beta_3 (S_W + S_H + S_V) \quad , \quad \beta_3 = \frac{4}{g_3^2} = \frac{4}{g^2(T)T} \tag{6}$$

with

$$S_W = \sum_{\mathbf{n}, i>j} \left(1 - \frac{1}{2} \operatorname{Tr} U_{ij}(\mathbf{n})\right) \tag{7}$$

being the three dimensional Wilson action,

$$S_H = -\sum_{\mathbf{n},i} \left(\frac{1}{2} \operatorname{Tr} D_i(U) A_4(\mathbf{n})\right)^2 \quad , \tag{8}$$

$$D_i(U)A_4(\mathbf{n}) = U_i(\mathbf{n})A_4(\mathbf{n}+\mathbf{e}_i)U_i^{\dagger}(\mathbf{n}) - A_4(\mathbf{n}) \quad ,$$

describing the coupling to the Higgs field, and $S_V$ being local Higgs self-interactions which are quadratic and quartic to their lowest orders, and carry additional powers of $g^2(T)$. For the Higgs field representing the temporal gauge field components, we have chosen the convention $A_4 = i\sum_{j=1}^{3} A_4^j \sigma_j$ where $\sigma_j$ denote the Pauli-matrices.

Basic properties of the gauge invariant correlation functions for spatial components of the gauge fields — e.g. the spatial Wilson loops — can be understood in terms of this effective theory. For instance, as this effective theory is confining, it is natural to expect that spatial Wilson loops obey an area law behaviour in the high temperature phase

$$W(R,S) = \langle \mathcal{P}[e^{i \oint_{R \times S} dx_\mu A_\mu}] \rangle \sim e^{-\sigma_s RS} \quad , \tag{9}$$

where the (ordered) path is over a rectangle of area $R \times S$, and $\sigma_s$ is called the spatial string tension, although one should stress that it is not related to properties of a physical potential in the (3+1)-dimensional theory. In the case of QCD the effective theory itself is quite complicated even at high temperatures, as the non-static modes do not decouple from the static sector.[7] An analysis of the temperature dependence of the spatial string tension, thus, yields information on the importance of the non-static sector for long-distance properties of high temperature QCD.

The existence of a non-vanishing spatial string tension, $\sigma_s$, in the high temperature phase of (3+1)-dimensional $SU(N)$ lattice gauge theory can be proven rigorously at finite lattice spacing.[8] However, despite its basic relevance for a better understanding of the non-perturbative structure of non-abelian gauge theories at high temperature, little effort has been undertaken to arrive at a quantitative description of the properties of the spatial string tension. The first numerical study[4] suggested that $\sigma_s$ stays non-zero, but temperature independent, in the high temperature phase of QCD. Some indications for an increase of $\sigma_s$ with temperature have been found recently.[9]

## 4. Measurements and Analysis

We present here the results of a detailed, high statistics analysis of the spatial string tension. Some of our results have previously been published in Ref. 10. The finite (and zero) temperature $SU(2)$ gauge theory has been simulated on lattices of size $32^3 \times N_\tau$, with $N_\tau$ ranging from 2 to 32 (Tab. 1). The simulations have been performed at two values of the gauge coupling, $\beta = 2.5115$ and $\beta = 2.74$, which correspond to the critical couplings for the deconfinement transition on lattices with temporal extent $N_{\tau,c} = 8$ and $N_{\tau,c} = 16$, respectively.[6] The lattice spacing thus changes by a factor $2.00 \pm 0.04$, where the error is caused by the uncertainty in both of the critical couplings.

In each case we have performed between 2000 and 5000 thermalization sweeps. Measurements were taken every 100 sweeps. The integrated autocorrelation times have been calculated on some representative operators. In all cases they turned out to be small compared to the above measurement rate. Besides the original data sets, we produced other sets by binning the data into blocks of two, and five measurements. The whole analysis was performed on the original data set as well as on the binned data. No systematic increase of errors with the block length was observed. We conclude that successive measurements can be treated as being effectively independent.

| $\beta$ | $N_\tau$ | $T/T_c$ | meas. | $S_{min}$ |
|---|---|---|---|---|
| 2.7400 | 32 | 0 | 835 | 4 |
| | 16 | 1 | 918 | 5 |
| | 12 | 1.33 | 720 | 4 |
| | 8 | 2 | 279 | 3 |
| | 6 | 2.67 | 477 | 3 |
| | 4 | 4 | 2111 | 3 |
| | 2 | 8 | 8582 | 3 |
| 2.5115 | 16 | 0 | 550 | 3 |
| | 8 | 1 | 1320 | 3 |
| | 6 | 1.33 | 2580 | 3 |
| | 4 | 2 | 1700 | 2 |
| | 2 | 4 | 9360 | 2 |

Table 1: The simulated lattice volumes $32^3 \times N_\tau$. The fourth column gives the number of gauge field configurations used in the analysis. $S_{min}$ labels the local potential $V_{T,S_{min}}$, used as an approximant for the asymptotic potential $V_T$ (Eq. 10).

Smeared on- and off-axis Wilson loops with the quark sources separated by multiples of the vectors $(1,0,0)$, $(1,1,0)$, and $(2,1,0)$ have been measured for the determination of the spatial string tension (correlator along $z$-axis). In addition to these separations we have choosen multiples of $(1,1,1)$, $(2,1,1)$, and $(2,2,1)$ for the zero temperature cases. In each direction the quarks have been separated by up to half of the lattice extent which gives distances up to $\frac{1}{\sqrt{2}}N_\sigma$ ($\frac{\sqrt{3}}{2}N_\sigma$) in the finite (zero) temperature case. We apply the smearing technique presented in Ref. 11 to the links orthogonal to the direction in which the correlator is taken ($z$ for the spatial loops, and $t$ for the "temporal" ones).

After 40 such smearing iterations (temperature dependent) *pseudo-potentials* are determined from the Wilson loops $W(\mathbf{R}, S)$ of shape $\mathbf{R} \times S$, with $\mathbf{R}$ pointing into one of the above directions:

$$V_T(\mathbf{R}) = \lim_{S \to \infty} V_{T,S}(\mathbf{R}) \quad , \qquad V_{T,S}(\mathbf{R}) = \ln \frac{W(\mathbf{R}, S)}{W(\mathbf{R}, S+1)} \quad . \tag{10}$$

Though, due to our smearing process, also temporal links are used for construction of the smeared Wilson loops, the (spatial) pseudo-potentials are not affected. This is due to the fact that the "3D-smeared" $qq$-creation operator as well as the corresponding "2D-smeared" operator share a nonvanishing overlap with all eigenstates of the $A_{1g}$ representation of the lattice symmetry group $D_{4h}$. Numerically, this has been checked for $N_\tau = 2$ at $\beta = 2.5115$. Because of faster convergence to maximal ground state overlap we have applied the 3D-smearing algorithm throughout our investigation.

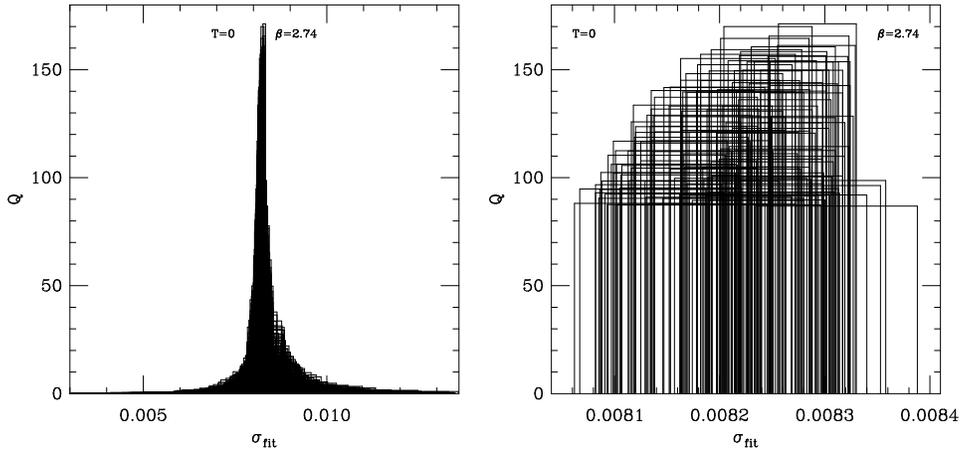

Figure 1: The quality $Q$ of various fits with different fit ranges $[R_{min}, R_{max}]$ against the fitted string tension $\kappa$ with error for $T = 0$, $\beta = 2.74$. In (a) all fits with confidence level $\alpha > 0.1$ are plotted. In (b) fits with quality $Q < Q_{max}/2$ are omitted. Note the different ordinate-scale.

We use $V_{T,S_{min}}(\mathbf{R})$ with finite $S_{min}$ as an approximant for the asymptotic value $V_T(\mathbf{R})$[†]. Our requirement is that the systematic error caused by this cutoff is small compared to the statistical error on $V_T$. To guarantee this we choose $S_{min}$ in the way that for all $S > S_{min}$ the individual potential values as well as the fitted value for the string tension remain constant within errors. The latter criterion is important because a small but systematic shift of potential values may have an enhanced influence on certain fit parameters.

The actual size of $S_{min}$ depends on statistics and ground state overlaps, as well as on the gap between ground state potential $V_T$ and the first excitation $V'_T$ in lattice units. Thus, for $\beta = 2.5115$ the asymptotics is reached earlier (in lattice units) than for $\beta = 2.74$. Also, for high temperatures not only the potential values are larger than in the zero temperature case but also the $V'_T(\mathbf{R}) - V_T(\mathbf{R})$ gaps seem to increase. So, $S_{min}$ decreases with increasing temperature, too, as can be seen from Tab. 1.

All our errors have been obtained from the scatter between (two times the number of measurements) *bootstrap* samples. The whole analysis has been performed within each bootstrap sample, separately. For the zero temperature cases on each bootstrap a different covariance matrix between potential values $V_T(\mathbf{R})$ has been calculated (with a "subbootstrap"). A comparison with the result obtained by using one and the same covariance matrix (obtained on the original data set) on all bootstrap channels did not show any significant change on errors. So, for the other fits the latter method was used for simplicity. All fits are correlated fits using this covariance matrix. We also cross-checked the results by comparing them with uncorrelated fits since correlated fits may lead to totally wrong results if the statistics is too small to fix the covariance matrix accurately, or the assumed parametrization does not completely match the data.[11]

---

[†] We prefer using "local masses" instead of exponential fits to Wilson loops with $S \geq S_{min}$ because of simplicity in data analysis. Due to the exponential increase of relative errors on pure gauge quantities with distance $S$, a fit would be governed by the values corresponding to $S_{min}$ and $S_{min} + 1$, anyway.

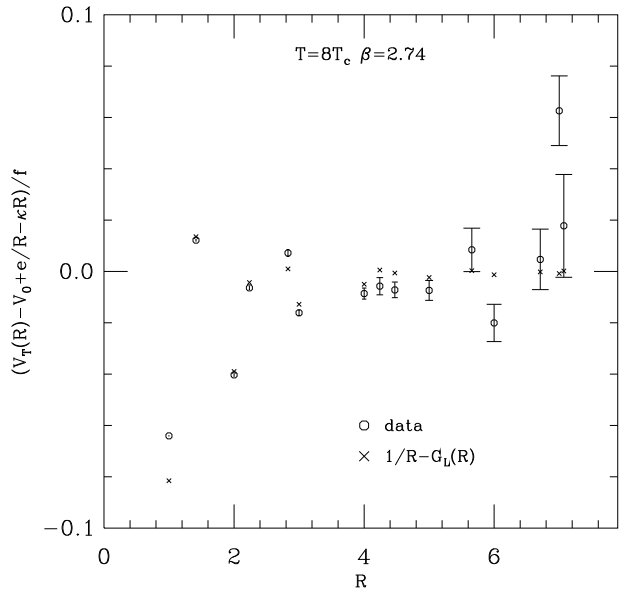

Figure 2: Comparison between deviations of the potential at $\beta = 2.74, T = 8T_c$ from the continuum symmetry parametrization $V(R) = V_0 + \kappa R - e/R$, and $1/R - G_L(\mathbf{R})$ (Eq. 12), which has been used to parametrize lattice artefacts.

The whole analysis has not only been performed for different $S_{min}$ on our original data sample and the two binned samples, but also with different fit ranges $[R_{min}, R_{max}]$. For all these fits (about 1000) we have calculated the confidence levels $\alpha = 1 - (2^{dof/2}\Gamma(dof/2))^{-1} \int_0^{\chi^2} dx\, x^{dof/2-1} e^{-x/2}$ and defined qualities $Q = \alpha \frac{dof}{dof_{max}} \frac{\kappa}{\Delta\kappa}$ where $\kappa$ are the fitted values for the string tension. The fit with highest quality gives the "best" fit range. In Fig. 1(a) a histogram of the quality $Q$ against the corresponding string tension value (with statistical error width) is plotted for all fits with $\alpha > 0.1$ for the zero temperature, $\beta = 2.74$ case. In Fig. 1(b) the same is plotted, but only for qualities $Q > Q_{max}/2$. We include the systematic error from the width of all the latter fits into all errors we state. As a check we also have defined a quality which uses the relative error of the Coulomb coefficient $e$ instead of the string tension $\kappa$. The resulting optimal fit ranges, as well as the systematic errors, have been almost the same in all cases.

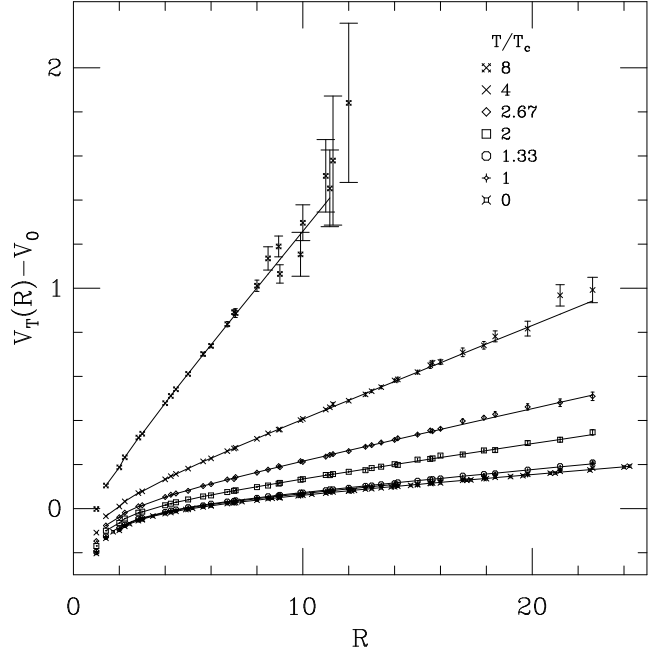

Figure 3: The pseudo-potentials $V_T(R)$ minus the (constant) self-energy contributions $V_0$ (Eq. 11) on lattices of size $32^3 \times N_\tau$ for $\beta = 2.74$ as a function of the spatial separation $R$, measured in lattice units.

| $\beta$ | $T/T_c$ | $V_0$ | $\kappa$ | $e$ | $f$ | $\sqrt{\sigma_s}/T_c$ |
|---|---|---|---|---|---|---|
| 2.7400 | 0 | .482 ( 3) | .0083 ( 1)( 1) | .220 (12) | .13 ( 8) | 1.46 (1)(1) |
| | 1 | .475 ( 6) | .0089 ( 2)( 6) | .210 (19) | .13 (12) | 1.51 (2)(5) |
| | 1.33 | .474 ( 3) | .0094 ( 1)( 2) | .207 ( 9) | .15 ( 6) | 1.55 (1)(2) |
| | 2 | .448 ( 5) | .0152 ( 1)( 5) | .175 (11) | .20 (11) | 1.97 (1)(3) |
| | 2.67 | .426 ( 6) | .0231 ( 2)( 5) | .157 (11) | .16 (10) | 2.43 (1)(3) |
| | 4 | .390 ( 4) | .0419 ( 2)( 4) | .135 ( 8) | .17 ( 7) | 3.28 (1)(2) |
| | 8 | .319 (11) | .1270 (11)(18) | .111 (17) | .28 ( 3) | 5.70 (2)(4) |
| 2.5115 | 0 | .537 ( 4) | .0337 ( 2)( 5) | .233 ( 8) | .26 ( 7) | 1.46 (1)(1) |
| | 1 | .543 ( 7) | .0325 ( 3)( 7) | .250 (16) | .20 (10) | 1.44 (1)(2) |
| | 1.33 | .513 ( 4) | .0381 ( 2)( 4) | .207 ( 7) | .24 ( 8) | 1.56 (1)(1) |
| | 2 | .443 ( 6) | .0643 ( 3)( 6) | .142 (13) | .27 ( 6) | 2.03 (1)(1) |
| | 4 | .350 (22) | .1715 (11)(32) | .128 (33) | .37 ( 8) | 3.31 (1)(3) |

Table 2: Summary of results from fits to the effective potentials using Eq. 11 on lattices of size $32^3 \times N_\tau$. The values of $N_\tau$ which correspond to the temperatures $T/T_c$ are given in Tab. 1. In the case of the string tension the first error is purely statistical. All other errors include systematic effects due to the choice of the fit range in the way described in the text.

## 5. Results and Discussion

### 5.1. *Evaluation of the spatial string tension*

In the way described above, several fits to various parametrizations have been performed. For a three parameter fit with self energy $V_0$, Coulomb coefficient $e$ and string tension $\kappa$ the $\chi^2$ values were found to be acceptable for $R_{min} \geq \sqrt{5}$. Fits turned out to be more stable (and $R_{min} \geq \sqrt{2}$) by use of the parametrization[13,14]

$$V_T(\mathbf{R}) = V_0 + \kappa R - \frac{e}{R} + f\left(\frac{1}{R} - G_L(\mathbf{R})\right) \qquad (11)$$

with

$$G_L(\mathbf{R}) = 4\pi \int_{-\pi}^{\pi} \frac{d^3k}{(2\pi)^3} \frac{\cos(\mathbf{kR})}{4\sum_i \sin^2(k_i/2)} \qquad (12)$$

being $4\pi$ times the lattice one-gluon-exchange for the infinite volume case. This last term takes account of the lattice artefacts present at small distances. The quality of this parametrization can be seen from Fig. 2 where a comparison (at $T = 8T_c$) between $1/R - G_L(\mathbf{R})$, calculated from the potential data by use of the fitted parameter values and the theoretical $1/R - G_L(\mathbf{R})$ is displayed.

We also have experimented with lattice propagators on smaller and anisotropic volumes, replacing the integral by a discrete sum over the allowed lattice momenta but, in general, the quality of fits decreased with smaller volume. We conclude that the potential values seem to be insensitive to the low momentum cutoff. Reasons for this might be that the Coulomb interaction from the mirror sources is screened by the large (non-perturbative) linear part of the potential and that Wilson loops have a well defined inside and outside.

Various other fit functions have been tried, including fits where the Coulomb part has been replaced by a logarithmic term, which would be expected in the three dimensional effective theory. We found that the fit parameter $\kappa = \sigma_s a^2$ only weakly depends on the actual parametrization of the short distance part of $V_T(\mathbf{R})$. Our results on the four fit parameters (Eq. 11) are summarized in Tab. 2. In Fig. 3 we display the potentials $V_T(R)$, measured at different temperatures, for the $\beta = 2.74$ case, together with the corresponding fit curves. We determine the spatial string tension in units of the critical temperature, $\sqrt{\sigma_s}/T_c = \sqrt{\kappa}N_{\tau,c}$, where $N_{\tau,c} = 8$ (16) for $\beta = 2.5115$ (2.74). These numbers are given in the last column of Tab. 2.

### 5.2. *Scaling behaviour*

A calculation of the "zero" temperature string tension in the confining phase on a

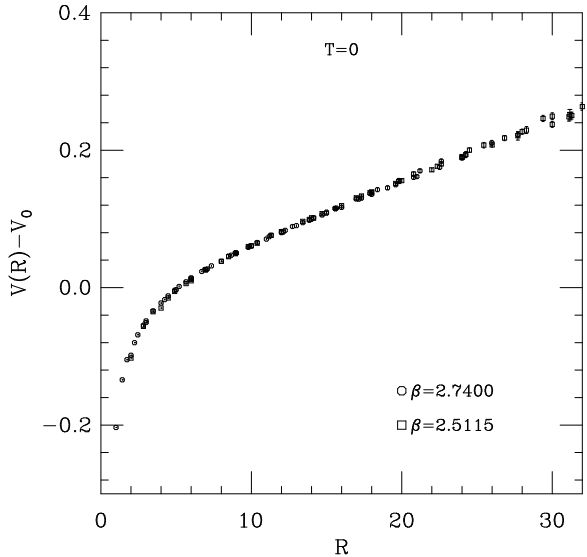

Figure 4: The physical zero temperature potentials minus the self energy parts $V_0$ (Eq. 11) at $\beta = 2.74$, and $\beta = 2.5115$ in units of the $\beta = 2.74$ lattice spacing. The $\beta = 2.5115$ values have been scaled by a factor $a_{\beta=2.5115}/a_{\beta=2.74} = 2.016$.

lattice of size $16 \times 32^3$ ($32^4$) at $\beta = 2.5115$ (2.74) yields

$$\sqrt{\sigma}a = \begin{cases} 0.1836 \pm 0.0013 & , \beta = 2.5115 \\ 0.0911 \pm 0.0008 & , \beta = 2.74 \end{cases} , \qquad (13)$$

which corresponds to a change in lattice spacing $a_{\beta=2.5115}/a_{\beta=2.74} = 2.016 \pm 0.023$, and is consistent with the factor $2.00 \pm 0.04$, obtained from the calculation of the critical couplings for the deconfinement transition.[6] This is visualized in Fig. 4 where the zero temperature potentials, measured at the two different $\beta$ values, are scaled by this factor 2.016.

In Fig. 5 we compare the spatial string tension calculated at $\beta = 2.5115$ and 2.74 at different temperatures. We find that — as in the zero temperature case — our data sets are consistent with each other down to a temporal extent as small as $N_\tau = 2$. Thus, scaling violations in the ratio $\sqrt{\sigma_s}/T_c$ are negligible. This demonstrates that the spatial string tension, indeed, is relevant to high temperature QCD as it persists in the continuum limit. Moreover, $\sigma_s$ coincides with the physical, zero temperature string tension for $T \leq T_c$.

5.3. *Comparison with the effective theory*

The coupling of the Yang-Mills part of the action of the effective 3-dimensional theory (Eq. 6), $g_3$, derived from a (3+1)-dimensional $SU(N)$ gauge theory at high temperature, is given in terms of the temperature and the four-dimensional coupling $g(T)$ as $g_3^2 = g^2(T)T$. Although the temperature will also set the scale for the Higgs self-couplings in the 3-dimensional theory, these couplings will in general have a different dependence on the four-dimensional gauge coupling $g(T)$.[15] The functional dependence of $\sigma_s(T)$ on $g^2(T)$ and $T$, thus, is not apparent from the general structure of the effective action.

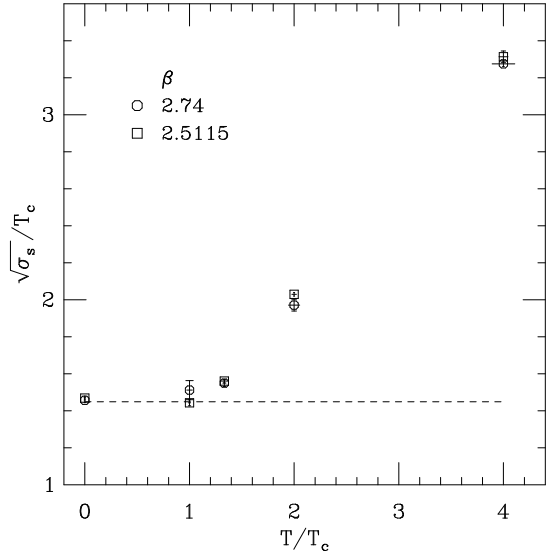

Figure 5: Square root of the spatial string tension in units of the critical temperature versus temperature, calculated at two different values of the gauge coupling. The broken line gives the result for the ratio of the physical string tension to the deconfinement temperature averaged over several values of the critical coupling.[6] The horizontal error bars indicate the uncertainty in the temperature scale $T_c$.

However, in a pure three-dimensional $SU(N)$ gauge theory dimensionful quantities are proportional to an appropriate power of the three-dimensional coupling $g_3$. If the temperature dependence of the pure gauge part of the effective action dominates the spatial string tension we would expect to find

$$\sqrt{\sigma_s(T)} = c g_3^2 = c g^2(T) T \quad , \tag{14}$$

where the temperature dependent running coupling constant $g^2(T)$ should, at high temperatures, be determined by the perturbative $\beta$-function of $SU(N)$ in four dimensions.

In Fig. 6 we have plotted $T/\sqrt{\sigma_s(T)}$ against $T$. From Eq. 14 this ratio is expected to be proportional to $g^{-2}(T)$. We have fitted these data to the two-loop formula for the coupling in $SU(2)$ gauge theory with the scale parameter $\Lambda_T$,

$$g^{-2}(T) = \frac{11}{12\pi^2} \ln T/\Lambda_T + \frac{17}{44\pi^2} \ln(2 \ln T/\Lambda_T) \quad . \tag{15}$$

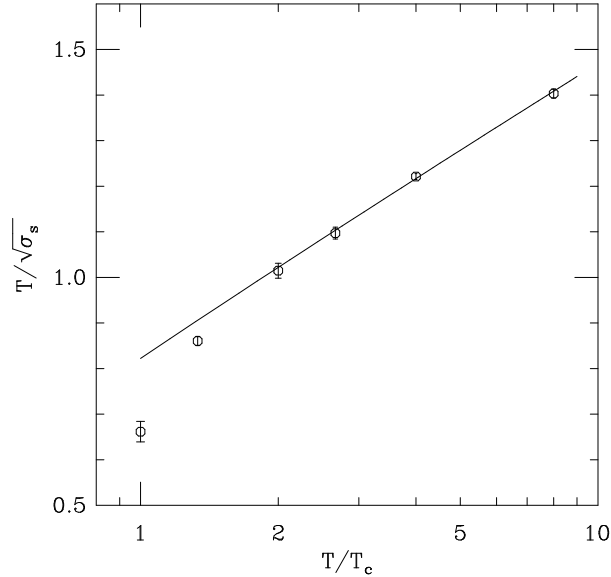

Figure 6: The ratio of the critical temperature to the square root of the spatial string tension versus temperature for $\beta = 2.74$. The line shows a fit to the data in the region $2 \leq T/T_c \leq 8$ using the 2-loop relation for $g(T)$ given in Eq. 15.

We find that the temperature dependence of the spatial string tension is well described by Eq. 14 and Eq. 15 for temperatures above $2T_c$. From the two parameter fit to the data shown in Fig. 6, we obtain

$$\sqrt{\sigma_s(T)} = (0.369 \pm 0.014) g^2(T) T \quad , \qquad (16)$$

with $\Lambda_T = 0.076(13) T_c$. We note that the second term in Eq. 15 varies only little with temperature. A fit with the one-loop formula thus works almost equally well; it yields $\Lambda_T^{(1)} = 0.050(10) T_c$ and $c = 0.334(14)$ for the coefficient in Eq. 16.

It is rather remarkable that the spatial string tension depends in this simple form on the perturbative $SU(2)$ $\beta$-function already for $T \geq 2T_c$. Moreover, we find that even quantitatively the spatial string tension agrees well with the string tension of the three-dimensional pure $SU(2)$ gauge theory,[16] $\sqrt{\sigma_3} = (0.3340 \pm 0.0025) g_3^2$.[‡] We take this as an indication that, indeed, the spatial string tension is dominated by the pure gauge part of the effective three-dimensional theory. There are also numerical indications[17] that a Higgs coupling of the strength suggested by perturbative arguments[5] has no significant influence on the potential values of three dimensional Yang-Mills theory. We note that the value for $g^2(T)$, determined here from long distance properties of the (3+1)-dimensional theory, is about a factor two larger than what has been obtained by comparing the short distance part of the (3+1)-dimensional heavy quark potential with perturbation theory.[5]

---
[‡]We have independently confirmed this value by use of smeared Wilson loops instead of Polyakov line correlators.[17]

Another quantity characterizing the high temperature behaviour is the electric Debye screening mass, $m_{el}$. It has been computed to one loop,[18] $m_{el} = \sqrt{2/3}\bar{g}(T)T$, with $\bar{g}(T)$ being a running coupling constant with scale parameter $\Lambda_{el}$. In one approximation this scale parameter has been computed to be[19] $\Lambda_{el} = 0.052\Lambda_{\overline{MS}}$, which can be compared to our result from the spatial string tension, $\Lambda_T^{(1)} = 0.062(14)\Lambda_{\overline{MS}}$ (here we have used[6] $T_c/\Lambda_{\overline{MS}} = 1.23 \pm 0.11$). It should be emphasized, though, that there is no unique definition of a temperature dependent coupling, $g(T)$, but it is tantalizing that $g(T)$ given by Eq. 16 might be quite close to $\bar{g}(T)$ determined from the Debye screening mass. We intend to investigate this possibility in the near future.

## Acknowledgements

The computations have been performed on the Connection Machines at HLRZ (16K CM-2, *fieldwise*), SCRI (16K partition of 64K CM-2, *slicewise*), and the Institut für Angewandte Informatik, Wuppertal (8K CM-2, *slicewise*). We thank the staff of these institutes for their support. The work of JF and UMH was supported in part by the DOE under grants # DE-FG05-85ER250000 and # DE-FG05-92ER40742. The work of GSB and KS was supported by the EC under grant # SC1*-CT91-0642. We are grateful to the DFG for supporting the Wuppertal CM-2 project (grant Schi 257/1-4).